\documentclass[aps,prd,superscriptaddress,floats,floatfix,preprint,nofootinbib]{revtex4-1}
\usepackage{amsmath}
\usepackage{amsfonts}
\usepackage{graphicx}
\usepackage{bbm}
\usepackage{multirow}
\usepackage{pifont}
\usepackage{hyperref}

\newcommand{\tbox}[1]{\mbox{\tiny #1}}

\newcommand{\WM}{\textit{Mathematica} }
\newcommand{\MW}{\textit{Mathematica} }

\newcommand{\RImporter}{\WM ROOT importer }
\newcommand{\MEight}{\textit{Mathematica} 8}
\newcommand{\MLink}{\textit{MathLink} }
\newcommand{\Import}{{\tt{Import[]}}}
\newcommand{\Export}{{\tt{Export[]}}}
\newcommand{\Courier}[1]{{\tt{#1}}}
\newcommand{\Cr}[1]{{\tt{#1}}}
\newcommand{\weblink}[1]{\href{#1}{#1}}

\begin{document}
\title{\WM with ROOT}

\author{Ken Hsieh}
\email[]{kenh@wolfram.com}
\affiliation{Wolfram Research, Inc., Champaign, IL 61820, USA}
\author{Thomas G. Throwe}
\affiliation{Brookhaven National Laboratory, Upton, NY 11973, USA}
\author{Sebastian White}
\affiliation{The Rockefeller University, NY, NY, 10065}

\preprint{}
\begin{abstract}
We present an open-source \textit{Mathematica} importer for CERN ROOT files.
Taking advantage of \textit{Mathematica}'s import/export plug-in mechanism, the importer offers a simple, unified interface
that cleanly wraps around its \MLink-based core that links the ROOT libraries with \textit{Mathematica}.
Among other tests for accuracy and efficiency, the importer has also been tested on a large (~5 Gbyte) file structure, D3PD, used by the ATLAS experiment for offline analysis without problems.
In addition to describing the installation and usage of the importer, we discuss how the importer may be
further improved and customized.
The package may be downloaded \href{http://library.wolfram.com/infocenter/Articles/7793/}{here} and a
related presentation may be found \href{http://cd-docdb.fnal.gov/cgi-bin/DisplayMeeting?conferenceid=522}{here}.
\end{abstract}

\maketitle

\section{Introduction}
\textit{Mathematica}\cite{Mathematica} and ROOT \cite{ROOT} are two powerful tools used in many technical fields.
In the field of high-energy physics, the different needs of theorists and experimentalists have traditionally
migrated the theorists towards Mathematica, while the experimentalists have relied on ROOT as one of the key
tools of analysis.
With the advent of the Large Hadron Collider (LHC), the need of collaboration between theorists and experimentalists is
as great as ever.

With this in mind, we present \RImporter.
Not only have we developed several functions able to import some data contained in ROOT files to \textit{Mathematica},
we also offer a simple and unified interface to use these functions, taking advantage of the new features of \Import and \Export of \MEight
\cite{MCreateImporter}.
While it does not capture all the possible rich data contained in many root files,
we present this program in the hope that it may be modified and tweaked to something more useful
not only to the HEP community, but to the broader \textit{Mathematica} and ROOT user bases.

It is important at the outset to list what the converter is, and is not, capable of.
As provided, it can
\begin{itemize}
\item import the list of objects (and their types) stored in a ROOT file through iteration of \Cr{TKey},
\item import the list of \Cr{TLeaf} stored in a \Cr{TTree}
\item import the data of single-leaf \Cr{TBranch} of the data type listed in Table \ref{tb:SupportedDataTypes}.
Essentially, we support only the basic C-types and some C++ standard template library containers containing these basic types.
\item import the bin and error data stored in \Cr{TH1F} and \Cr{TH2F} objects and create histograms from the bin data.
\end{itemize}
Some of the shortcomings include
\begin{itemize}
\item import data from \Cr{TBranch}es that contain multiple \Cr{TLeaf}s,
\item import data from \Cr{TBranch}es whose data type are not listed in Table~\ref{tb:SupportedDataTypes},
\item import other types of histogram \Cr{TH1*} and \Cr{TH2*} objects.
\end{itemize}
However, it should be noted that in some cases it is relatively straightforward to add additional data
and histogram types.

We include a few example data files but this package has also been tested on a large (~5 Gbyte) file structure, D3PD, used by the ATLAS experiment
for offline analysis without problems.
	
In this initial release we are only distributing source files with executables for the Windows platform.
For Mac and Linux machines, we include several makefiles for creating executables on the different platforms
rather than the executables themselves.
	
\begin{table}[h]
\begin{center}
\begin{tabular}{|l|c|c|c|c|}
\hline
Type (\Cr{T}) & \Cr{   T   } & \Cr{v<T>} & \Cr{v<v<T> >} & \Cr{l<T>}  \\
\hline
\Cr{int} & x & x & x & x \\
\hline
\Cr{unsigned int} & x & x & x & x \\
\hline
\Cr{short} & x & x & x & x \\
\hline
\Cr{unsigned short} & x & x & x & x \\
\hline
\Cr{double} & x & x &x& x \\
\hline
\Cr{float} & x & x & x & x \\
\hline
\Cr{bool} & x & x & x &x \\
\hline
\Cr{string} & x & x & x & x \\
\hline
\Cr{char*} & x & &  & \\
\hline
\end{tabular}
\label{tb:SupportedDataTypes}
\caption{
Supported \Cr{TBranch} data types.
The abbreviation \Cr{v<T>} and \Cr{l<t>} stand respectively for \Cr{std::vector<T>} and \Cr{std::list<T>},
where \Cr{T} is a generic data type.
The \Cr{string} listed here is \Cr{std::string} of \Cr{C++}.
In some cases of \Cr{std::vector<std::vector<T> >} and \Cr{std::list<T>}, libraries may need to be generated.
}
\end{center}
\end{table}

\section{Installation}
\subsection{Package Contents}
The Mathematica ROOT importer package can be downloaded at
\newline \weblink{http://library.wolfram.com/infocenter/Articles/7793/}.
In addition to the folder \Cr{References} containing several references and guides, the package
contains the following files:
\begin{center}
\begin{tabular}{ll}
\Cr{        }&\Cr{Examples/Mathematica\_ROOT\_M8\_Usage.nb}\\
\Cr{        }&\Cr{Examples/Mathematica\_ROOT\_Tests.nb}\\
\Cr{        }&\Cr{Examples/basic.root}\\
\Cr{        }&\Cr{Examples/cernstaff.root}\\
\Cr{        }&\Cr{Examples/demo.root}\\
\Cr{        }&\Cr{Examples/l1.root}\\
\Cr{        }&\Cr{Examples/s1.root}\\
\Cr{        }&\Cr{Examples/th2f.root}\\
\Cr{        }&\Cr{Examples/v1.root}\\
\Cr{        }&\Cr{Examples/v2.root}\\
\Cr{        }&\Cr{ROOT/BuildMathLinkExecutable.nb}\\
\Cr{        }&\Cr{ROOT/Converter.m}\\
\Cr{        }&\Cr{ROOT/Import.m}\\
\Cr{        }&\Cr{ROOT/makefile}\\
\Cr{        }&\Cr{ROOT/makefile.linux32}\\
\Cr{        }&\Cr{ROOT/makefile.linux64}\\
\Cr{        }&\Cr{ROOT/makefile.mac32}\\
\Cr{        }&\Cr{ROOT/makefile.mac64}\\
\Cr{        }&\Cr{ROOT/root\_interface.tm}\\
\Cr{        }&\Cr{ROOT/ROOT.sln}\\
\Cr{        }&\Cr{ROOT/ROOT.vcproj}\\
\Cr{        }&\Cr{ROOT/Binaries/*/}\\
\Cr{        }&\Cr{ROOT/Binaries/[Windows,Windows-x86-64]/ROOT.exe}\\
\end{tabular}
\end{center}
where * spans those architectures supported by \MEight
\Cr{\{Linux, Linux-x86-64, MacOSX-x86, MacOSX-x86-64, Windows, Windows-x86-64\}}
.

\subsection{Requirements}
The \textit{Mathematica} ROOT importer requires \textit{Mathematica} 8 and
CERN ROOT\footnote{The \RImporter is developed and tested with ROOT
5.28/00.}.
The \MLink portion of the importer dynamically links the ROOT libraries
at compile-time, and loads the libraries at run-time.
At compile-time (only needed on Linux and Mac machines),
the users have to supply the paths to the ROOT header and libraries
files to \Cr{ROOT/BuildMathLinkExecutable.nb} or the appropriate \Cr{makefile}
in order to compile successfully
During run-time, the \MLink executable and the ROOT libraries (which
may, in turn, load other ROOT libraries)
depend on environment variables to locate the ROOT libraries.

On Windows machines, the path to the ROOT libraries should be included in
the environment variables \Cr{\$Path} and \Cr{\$LIB}.
On Mac and Linux machine, if the path to the ROOT libraries is not explicitly compiled into the
ROOT binaries and libraries, then, at run-time, the location of the libraries is made available through an environment variable.
The user may check whether the ROOT library path is compiled into the binaries
and libraries by running the \Cr{ldd} (Linux) or \Cr{otool -L} (Mac)
command on the appropriate binaries and libraries.  If these commands
indicate that the location of the libraries is unknown, then the
\Cr{\$LD\_LIBRARY\_PATH} (Linux) and \Cr{\$DYLD\_LIBRARY\_PATH} (Mac)
are used to indicate the location.
The environment variable is generally set or appended in the user's
\Cr{.bash\_profile} or equivalent file for the user's particular shell.
Depending on the user's operating system and user interface, environment
variables may or may not be available to applications launched through a
menu or desktop shortcut.  If the path to the ROOT libraries is not
compiled into the binaries and libraries, and if evaluating
\Cr{Environment["LD\_LIBRARY\_PATH"]} returns \Cr{\$Failed} when
\textit{Mathematica} is launched in this way, then the user will be
required to start \textit{Mathematica} from a terminal session in order
to have access to the environment and use the importer.

%

\subsection{Installation}
The installation process is divided into two main steps: generating the \MLink executable and
copying the necessary files to a location where \WM may load it automatically.

\subsubsection{Generating the \MLink executable}
This step is needed only for users on Mac and Linux machines.
For Windows users, we include pre-compiled \MLink executables for Windows platform (built with ROOT 5.28/00)
and the users may skip directly to the next step.

A \MLink executable needs to compiled from the file \Cr{ROOT/root\_interface.tm} included in the package.
The compilation process requires first processing the \Cr{.tm} file into a \Cr{.cpp} file using
the \WM utility \Cr{mprep},
and then compiling the resultant \Cr{.cpp} while linking against the ROOT libraries.
We have included a Visual Studio project file for the Windows platform and a makefile for the Linux and Mac
platforms to build the executables.
One would have to modify a few items (such as a location of the local ROOT libraries) before being able
to compile successfully.
The executables can also be generated using \WM with the function \Cr{CreateExecutable}, and our implementation
to build the executable using \Cr{CreateExecutable} \cite{MCreateExecutable} is included in \Cr{ROOT/BuildMathLinkExecutable.nb}.

The compilation target is a \MLink executable named \Cr{ROOT.exe}, and it should be placed appropriately,
depending on the platform, inside one of the folders in \Cr{ROOT/Binaries/} as set up the makefiles.
For example, the makefile for on 64-bit Linux machines would place \Cr{ROOT.exe} under the
folder \Cr{ROOT/Binaries/Linux-x86-64/}.

\subsubsection{Copy files to \Cr{\$UserBaseDirectory/SystemFiles/Formats} }
To install the \RImporter, simply copy the content of the \Cr{ROOT} folder into the directory
\Cr{\$UserBaseDirectory/SystemFiles/Formats}, where the \WM  path variable \Cr{\$UserBaseDirectory}
can be found by evaluating \Cr{\$UserBaseDirectory} in \WM.  (In some cases, it may be necessary to create
the folder \Cr{Formats} under \Cr{\$UserBaseDirectory/SystemFiles}.)
Similarly, the converter may be uninstalled by deleting this copy of the \Cr{ROOT} folder.

\subsection{Library-Generation during first-time use}
When the \RImporter is used for the first time, it generates the libraries needed
by ROOT to support \Cr{vector<vector<*> >} types.
The libraries are generated in the directory \Cr{\$UserBaseDirectory/SystemFiles/Formats/ROOT} along with
a lock file \Cr{CreateLibrary.m}, whose presence signals to \RImporter not to generate the libraries
again in a new session.

When the user upgrades ROOT and needs to compile a new version of the \MLink executable,
the libraries would also need to be generated.  This can be done by the new \MLink executable
once the libraries (files starting with the prefix \Cr{AutoDict*}) and the lock file
\Cr{CreateLibrary.m} are removed.

\section{Usage}
Once the ROOT import package is installed, there are several usages depending on the type of object
stored in the ROOT file and the scope of information requested by the user.
It is recommended that the users are familiar with the general syntax of the \WM \Import function \cite{MImport}.
We list the possible usage in Table \ref{tb:AllImportCalls} before explaining each one in detail.
It is important to note several things:
\begin{itemize}
\item Since \Cr{"ROOT"} is a user-defined format, we must explicitly denote the format name.
\item Except for the \Cr{"Keys"} element, all the other elements require a sub-element: the name of the object to be inspected/imported.
\item The returned structure and the options available to each element differ, and are explained in detail in
the following subsections.
\item The table only lists the native elements defined by the ROOT converter.
The general, format-independent features of the \WM \Import/\Export framework (for example: importing
multiple elements at once, the \Cr{"Elements"} element, etc.) are not listed here.
\end{itemize}

\begin{table}[h]
\begin{center}
\caption{
The element names and brief description of the \MW ROOT importer.
The items in the left column are understood to be used as the second argument
to the \MW \Import function (with the first argument being the file name).
The italicized keywords are the user-supplied variables,
where \textit{file} is the name of the imported ROOT file.
The variable \textit{tree} is the name of a \Cr{TTree} object in the ROOT file \textit{file}
and \textit{branch} is the name of a \Cr{TBranch} object in \textit{tree}.
The variable \textit{hist} is the name of a \Cr{TH1F} or \Cr{TH2F} object.}
\label{tb:AllImportCalls}
\vspace{0.125in}
\begin{tabular}{|l|l|}
\hline
\Cr{Import[}\textit{file}\Cr{, \_\_\_  ]} & Description \\
\hline
\Cr{\{"ROOT", "Keys"\}}, &
\Cr{TKey} information, such as class names. \\
\hline
\Cr{\{"ROOT", "TTreeMetadata",}\textit{tree} \Cr{\}} &
\Cr{TTree} meta-info., such as \Cr{TBranch} names and data types \\
\hline
\Cr{\{"ROOT", "TTreeData",}\textit{tree}\Cr{\}} &
data from all \Cr{TBranch}'s contained in \Cr{TTree} \textit{tree} \\
\hline
\Cr{\{"ROOT", "TTreeData",}\textit{tree}\Cr{,}\textit{branch}\Cr{\}} &
data from a particular \Cr{TBranch} \textit{branch} contained in \Cr{TTree} \textit{tree} \\
\hline
\Cr{\{"ROOT", "TH1FData",}\textit{hist}\Cr{\}} &
data in a \Cr{TH1F} object given as a formatted list of numbers \\
\hline
\Cr{\{"ROOT", "TH1FGraphics",}\textit{hist}\Cr{\}} &
\Cr{TH1F} object rendered using the \Cr{Histogram[]} \\
\hline
\Cr{\{"ROOT", "TH2FData",}\textit{hist}\Cr{\}} &
data in a \Cr{TH2F} object given as a formatted list of numbers \\
\hline
\Cr{\{"ROOT", "TH2FGraphics",}\textit{hist}\Cr{\}} &
\Cr{TH2F} object rendered using the \Cr{Histogram3D[]} \\
\hline
\end{tabular}
\end{center}
\end{table}
%

\subsection{Inspect of contents of a ROOT file through its keys}
To inspect the content of a ROOT file, we may import its \Courier{"Keys"} element:
\begin{center}
\Cr{Import[}\textit{file}\Cr{, \{"ROOT", "Keys"\}]},
\end{center}
and since \Cr{"Keys"} is the default element for the \Cr{"ROOT"} format,
we can equivalently specify only the format and drop the head \Cr{List}:
\begin{center}
\Cr{Import[}\textit{file}\Cr{, "ROOT"]}.
\end{center}%
The output of the function call is a list of \Cr{"TKey"} information \{$\mbox{KeyInfo}_1$, $\mbox{KeyInfo}_2$, ...\},
with each entry is itself a triplet:
\begin{eqnarray}
\mbox{KeyInfo}_i = \{ \mbox{Key Name}_i, \mbox{Key Title}_i, \mbox{Class Name}_i \}.
\end{eqnarray}
Some examples of inspecting ROOT files using the \Cr{"Keys"} element is given in Fig. 1.

\begin{figure}
  \includegraphics[height=.24\textheight]{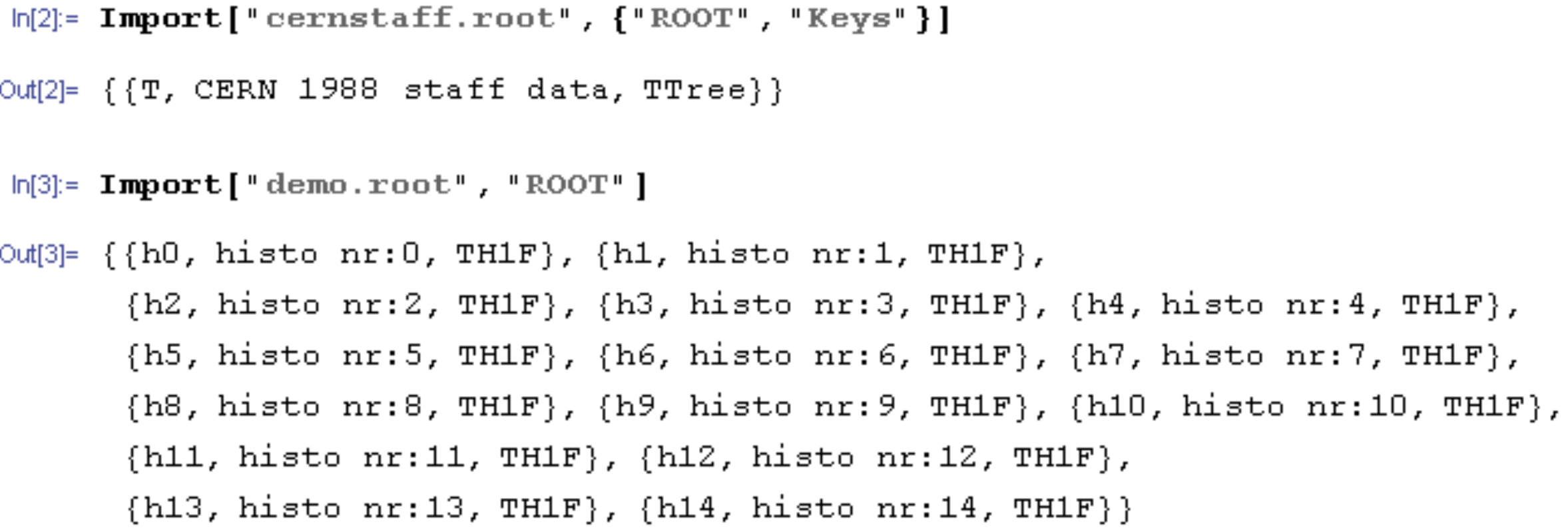}
  \caption{Examples of importing ROOT data with the \Cr{"Keys"} element.
In these examples, we see that the file \Cr{cernstaff.root} contains a
lone \Cr{TTree} object named \Cr{"T"} while the file \Cr{demo.root}
contains a collection of \Cr{TH1F} objects with names of the form \Cr{"h}$i$\Cr{"}.}
\end{figure}

\subsection{Inspect the branch information of a \Courier{TTree} object}
The data in a ROOT file is typically stored and organized in \Cr{TTree} objects, and
we may use the \Cr{"TTreeMetadata"} element to import the metainformation of a \Cr{TTree} object.
The function call
\begin{center}
\Cr{Import[}\textit{file}\Cr{, \{"ROOT", "TTreeMetadata",} \textit{tree} \Cr{]}.
\end{center}
The output is a list of quartets of the form:
\begin{eqnarray}
\{
\{ \mbox{branch name}_1, \mbox{branch title}_1,  \mbox{data type}_1, N_T \},
\{ \mbox{branch name}_2, \mbox{branch title}_2,  \mbox{data type}_2, N_T \}, ...
\}
\end{eqnarray}
where $N_T$ is the number of data entries of each branch, which should be the same throughout a particular
\Cr{TTree}.
An example of importing \Cr{TTree} metainformation is given in Fig.~2.

\begin{figure}
  \includegraphics[height=.17\textheight]{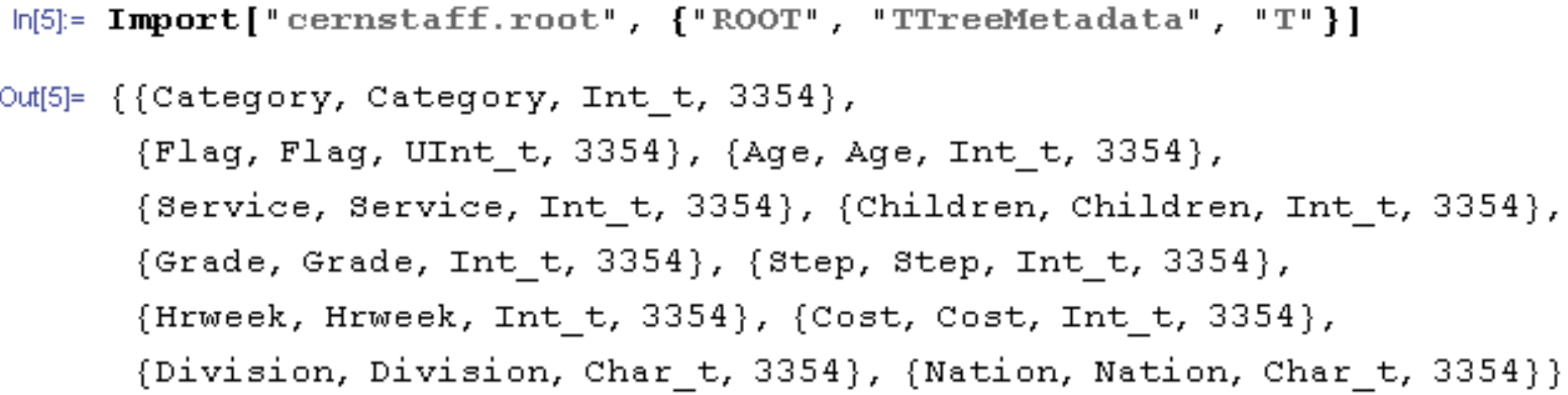}
  \caption{Example of importing ROOT data with the \Cr{"TTreeMetadata"} element.
In this example, we see that the \Cr{TTree} object named \Cr{"T"} in the file \Cr{cernstaff.root} contains
a collection of \Cr{TBranch}'s.  The first \Cr{TBranch} has name \Cr{"Category"} (the first item in the
list) with a title (second item) same as its name.  It stores data in the form of 3354 \Cr{Int\_t} objects.}
\end{figure}

\subsection{Import a particular \Courier{TBranch} object}
Knowing the name of a \Cr{TBranch} object, we can proceed to import the data inside the branch.
We may do this via the import element \Cr{"TTreeData"}:
\begin{center}
\Cr{Import[}\textit{file}\Cr{,\{"ROOT","TTreeData",} \textit{tree}\Cr{,}\textit{branch}\Cr{]}.
\end{center}%
and \WM returns the data stored in the \Cr{TBranch} object named \textit{branch}.

For large data sets, it may be particularly useful to import only parts of a branch.
This may be accomplished with the \Cr{"Range"} option with the value in the form of $\{m, n\}$,
with $m$ and $n$ both being positive integers:
\begin{center}
\Cr{Import[}\textit{file}\Cr{,\{"ROOT","TTreeData",} \textit{tree}\Cr{,}\textit{branch}
\Cr{\},"Range"}$\rightarrow$\Cr{\{}$m$\Cr{,}$n$\Cr{\}]}.
\end{center}
In this case, \WM imports only the $m^{\tbox{th}}$ through $n^{\tbox{th}}$ entries, inclusively.
Examples of importing data from \Cr{TBranch} objects are shown in Fig.~3.

\begin{figure}
  \includegraphics[height=.22\textheight]{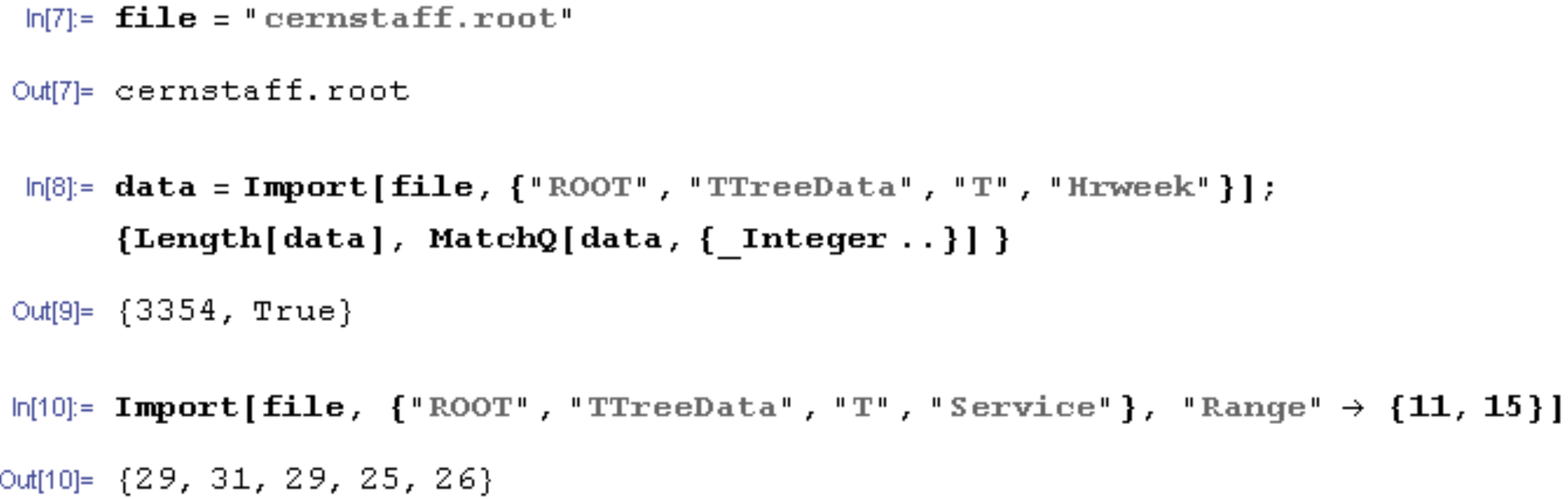}
  \caption{Example of importing ROOT data with the \Cr{"TTreeData"} element.
In the first example, we stored the data in the \Cr{TBranch} named \Cr{"Hrweek"}.
As the data is too large to be shown here, we merely show that we indeed have a collection
of integers.
The second example uses the \Cr{"Range"} option to import only the 11th through 15th entries
in the \Cr{TBranch} named \Cr{"Service"}}
\end{figure}

This feature, full and partial import of the data in a \Cr{TBranch} is arguably
the most important feature of the converter.
However, it is limited to \Cr{TBranch}'s that contain basic types of data, listed in Table \ref{tb:SupportedDataTypes}.
The users may extend the converter to work with additional types of data, as discussed in a later section.

\subsection{Import all branches of a \Courier{TTree} object}
If we import the \Cr{"TTreeData} without specifying a \Cr{TBranch} name, the importer
will iterate through the list of \Cr{TBranch} names obtained via \Cr{"TTreeMetadata"}.
In addition, the \WM \Import/\Export framework automatically parses its arguments
and we may import several \Cr{TBranch}'s (as long as they belong to the same \Cr{TTree})
in a single \Import call.
As with the specific \Cr{TBranch} importer, the full \Cr{TTree} importer also takes
a \Cr{"Range"} option, and imports only specified entries for all the \Cr{TBranch}'s.
These features are illustrated in Fig.~4.

\begin{figure}
  \includegraphics[height=.32\textheight]{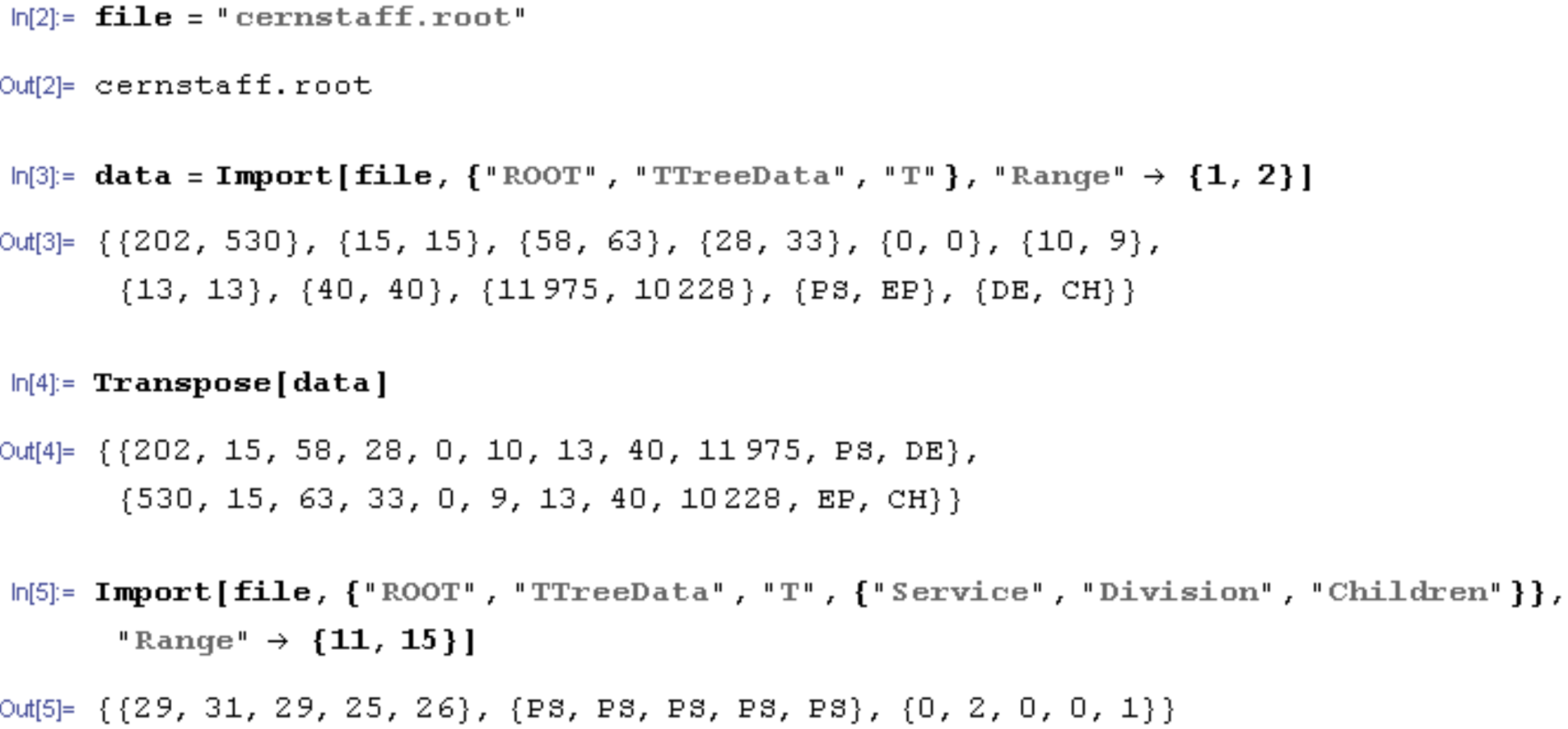}
  \caption{Examples of importing multiple \Cr{TBranch}'s with the \Cr{"TTreeData"} element.
In the first example, we import the first two entries of each branch in \Cr{TTree} \Cr{"T"}.
(Without the \Cr{"Range"} option, we would have imported all the entries.)
It is often desirable to \Cr{Transpose[]} the result, giving us the first two records
in \Cr{"T"} (note that the order of the elements are the same as that given by
the \Cr{"TTreeMetadata"} element.
In the last example, we import some specific entries from three selected
\Cr{TBranch}'s.
}
\end{figure}

\subsection{Import the data of a \Courier{TH1F} object}
To import the data contained in the histogram, we use the \Cr{"TH1FData"} element:
\begin{center}
\Cr{Import[}\textit{file}\Cr{,\{"ROOT","TH1FData",} \textit{hist}\Cr{\}]}.
\end{center}%
The function returns a list of bin data, with entry in the form of
\begin{eqnarray}
\{ x_i, \Delta x_i, c_i, \Delta c_i \},
\end{eqnarray}
where $x_i$ and $\Delta x_i$ are respectively the lower edge and the width of bin $i$,
and $c_i$ and $\Delta c_i$ are respectively bin content and its error.

\subsection{Import the data of a \Courier{TH1F} object as a histogram}
We can import the \Cr{TH1F} object directly as a \WM graphics via the \Cr{"TH1FGraphics"} element:
\begin{center}
\Cr{Import[}\textit{file}\Cr{,\{"ROOT","TH1FGraphics",} \textit{hist}\Cr{\}]}.
\end{center}
The function directly returns the graphics rendered using the \Cr{Histogram} function of \WM.

The \Cr{"TH1FGraphics"} also takes as options those options available to \Cr{Histogram}.
The list of option names and their default values can be retrieved by evaluating
\Cr{Options[Histogram]}.

\subsection{Import the data of a \Courier{TH2F} object}
To import the data contained in a two-dimensional histogram, we use the \Cr{"TH2FData"} element:
\begin{center}
\Cr{Import[}\textit{file}\Cr{,\{"ROOT","TH2FData",} \textit{hist}\Cr{\}]}.
\end{center}%
The function returns a list of bin data, with entry in the form of
\begin{eqnarray}
\{ x_i, \Delta x_i, y_i, \Delta y_i,c_i, \Delta c_i \},
\end{eqnarray}
where $\{x_i, y_i\}$ and $\{\Delta x_i, \Delta y_i\}$ are respectively
the coordinate of the lower edge and widths of bin $i$,
and $c_i$ and $\Delta c_i$ are respectively bin content and its error.

\subsection{Import the data of a \Courier{TH2F} object as a histogram}
We can import the \Cr{TH2F} object directly as a \WM graphics via the \Cr{"TH2FGraphics"} element:
\begin{center}
\Cr{Import[}\textit{file}\Cr{,\{"ROOT","TH2FGraphics",} \textit{hist}\Cr{\}]}.
\end{center}
The function directly returns the graphics rendered using the \Cr{Histogram3D} function of \WM.

The \Cr{"TH2FGraphics"} also takes as options those options available to \Cr{Histogram3D}.
The list of option names and their default values can be retrieved by evaluating
\Cr{Options[Histogram3D]}.

\section{Brief Remarks about the Internal Workings and Extensibility}
In this section we offer several remarks about the internal workings of
the \WM ROOT importer that may be of interest to those users interested in
extending the converter.

\subsection{Internal workings}
The \WM ROOT importer is based on three technologies: CERN ROOT to extract
the data inside a ROOT file, \MLink to transmit the extracted data to Mathematica,
and using \WM to post-process the data as necessary.
The ROOT and \MLink portions of the code are contained in the file \Cr{root\_interface.tm},
and this \MLink template file needs to be processed by the \Cr{mprep} utility included
in \WM before it can be compiled using a C++ compiler.
The \MLink executable defines several low-level functions defined under the
\Cr{ROOTImport`} context.
The input signatures of these functions are defined in \Cr{root\_interface.tm}.
For example,
\begin{center}
\Cr{ROOTImport`importKey["/home/kenh/Documents/ROOTConvert/Examples/cernstaff.root"]}
\end{center}
returns the same output as \Cr{Import["cernstaff.root",\{"ROOT","Keys"\}]} because
the "Keys" importer essentially wraps around the low-level function.
There are, however, several importance advantages using the \Import call:
\begin{itemize}
\item the \Import framework automatically passes the full path to the converter functions,
\item among other checks, the \Import framework checks that the file actually exists before calling the low-level functions
\item the \Import framework provides an easy-to-use interface that may call several low-level functions in one input and allows
users to customize the form of the output
\end{itemize}

\subsection{Extensibility}
Our current implementation likely needs to be extended before reaching its full potential
as a bridge between Mathematica and ROOT.
Recognizing this, we have taken the effort to document both the \MLink and \WM portions of the code,
and the existing functions may serve as templates for further development.

Occasionally, the authors may also provide fixes and new features to the code, in which case the online
version of this document will be correspondingly updated.

\section{Conclusion}
This paper presents a simple implementation of a \WM importer for ROOT that is able to import some
data stored in \Cr{TTree} and the histogram objects \Cr{TH1F} and \Cr{TH2F}.
Taking advantage of the import/export plug-in mechanism, the importer offers an easy-to-use interface
while the core
While it does not capture all the possible rich and flexible
data types typically stored in the ROOT files, it is nonetheless an effective and useful first step,
as evident in our usage/test in importing a large ATLAS data set.
The open-source nature of the project opens doors for the importer to be improved and customized,
and we hope this tool will be useful to broad communities of \WM and ROOT users.

\begin{acknowledgments}
We are grateful to Stephen Wolfram and Peter Overmann for dedicating the resources that made this project possible.
We also thank Philippe Canal, Valeri Fine, Pavel Nevski, and Fons Rademakers for helping us with various aspects of CERN ROOT.
KH would also like to thank Abdul Dakkak and William Sehorn for assisting with parts of this project.
\end{acknowledgments}


\end{document}